\documentclass{DISproc}

\newcommand{\xB}{x_{\rm B}}
\newcommand{\Q}{{\cal Q}}
\newcommand{\KMa}{{\it KM10a }}
\newcommand{\KMb}{{\it KM10b }}
\newcommand{\KM}{{\it KM10 }}
\newcommand{\GK}{{\it GK07 }}

\begin{document}

\title{GPDs from present and future measurements}

\author{{\slshape Kre\v{s}imir Kumeri{\v c}ki$^1$ and Dieter M\"uller$^2$\footnote{Presented by D.M. at the XX International Workshop on 
Deep-Inelastic Scattering and Related Subjects, 26-30 March 2012, University of Bonn.}}\\[1ex]
$^1$Department of Physics, University of Zagreb, Zagreb, HR-10002, Croatia\\
$^2$Ruhr-University Bochum, Bochum, D-44780, Germany}

\maketitle



\vspace{-4mm}
\begin{abstract}
We report on the access of generalized parton distributions (GPDs) from deeply virtual Compton scattering (DVCS) measurements. We also point out that such measurements at a proposed high-luminosity electron-ion collider (EIC) provide insight in both the transverse distribution of sea quarks and gluons as well as the proton spin decomposition.
\end{abstract}

\vspace{-4mm}
\section{Introduction}

During the last decade the HERA  and JLAB collaborations have spent much effort
to measure exclusive processes such as electroproduction of photon, vector
mesons, and pseudoscalar mesons in the deeply virtual region in which the
virtuality $\Q^2 \gtrsim 1 \GeV^2$ of the exchanged space-like photon allows to
resolve the internal structure of the proton. In such processes one can access
GPDs which can be viewed as a non-diagonal overlap of light-cone wave
functions. These GPDs are intricate functions $F(x,\eta,t,\mu^2)$ that depend
on the momentum fraction $x$, the skewness $\eta$, the $t$-channel momentum
transfer $t$, and the factorization scale $\mu^2$.
Phenomenologically, they are
most directly accessible at the crossover line $\eta=x$ (see below).  Moreover,
GPDs are related to (generalized) form factors and standard parton densities.
The GPD framework opens a complementary window to address the partonic content
of the nucleon. In particular, it offers the possibility to access  the
transverse distribution of partons  and to address the decomposition of the
nucleon spin in terms of quark and gluon degrees of freedom, see reviews
\cite{Die03aBelRad05}.

In Sect.~2 we shortly report on the GPD extraction from present DVCS data and in Sect.~3 we study the possible impact of an proposed EIC. Finally, we conclude.

\vspace{-2mm}
\section{GPDs from present DVCS experiments}

\begin{figure}[t]
\centering
\vspace{-4mm}
\includegraphics[width=1\textwidth]{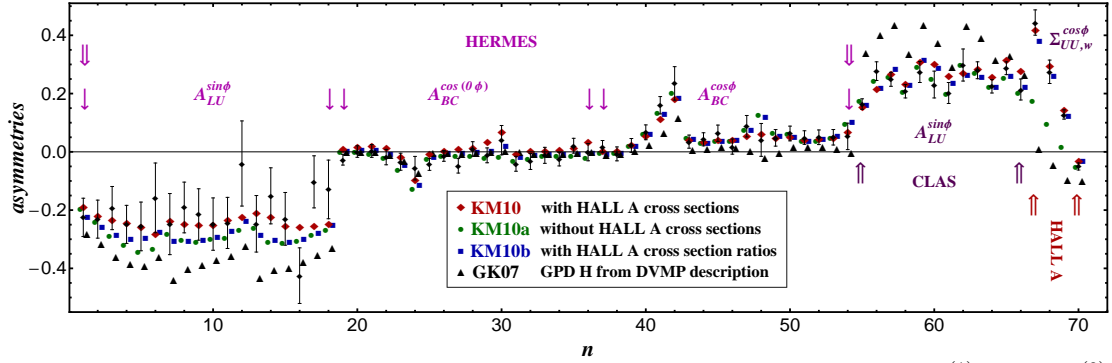}
\vspace{-8mm}
\caption{\small
Measurements for fixed target kinematics labeled by data point number $n$: $A^{(1)}_{\rm LU}$ (1-18),  $A^{(0)}_{\rm BC}$  (19-36), $A^{(1)}_{\rm BC}$ (37-54) from \cite{Airapetian:2009aa}; $A^{(1)}_{\rm LU}$  (55-66)
and $\Sigma_{\rm LU}^{(1),w}$  (67-70) are derived from Refs.~\cite{Girod:2007aa}
and \cite{Munoz_Camacho:2006hx}. Hybrid model fits \KMa (circles, slightly shifted to the l.h.s.) \KMb  (squares, slightly shifted to the r.h.s.) and {\KM} (diamonds) and a model prediction \GK (triangles-up, slightly shifted to the r.h.s.) \cite{Goloskokov:2007nt}.
}
\label{Fig1}
\end{figure}
The DVCS amplitude is mostly dominated by the Compton form factor (CFF) ${\cal
H}$ and to extract information on the associated GPD $H$ we consider DVCS off
an unpolarized proton target. The leading twist-two dominated beam spin $A_{\rm
LU}^{\sin\phi}$, beam charge $A_{\rm BC}^{\cos\phi}$ and $A_{\rm
BC}^{\cos(0\phi)}$ as well as azimuthal angle $\Sigma_{{\rm UU},w}^{\cos\phi}$
asymmetry data  from fixed target experiments together with fit results
\cite{Kumericki:2009uqKumericki:2011zc} and model predictions
\cite{Goloskokov:2007nt} are displayed in Fig.~\ref{Fig1}. Here,  $\mbox{\it
KM10}_{\ldots}$ arise from DVCS fits that include also H1/ZEUS collider data
and utilize hybrid models where sea quarks and gluons are described
by flexible GPD models while the valence quark GPDs are only modeled on the
crossover line and dispersion relations are used for the later to evaluate the
corresponding real part of the CFFs \cite{Kumericki:2009uqKumericki:2011zc}.
It turns out that the unpolarized cross section measurements of Hall A at rather
large $\xB=0.36$ indicate a larger DVCS amplitude,
which is not expected from ``standard'' GPD models. In
the \KMa fit we simply neglect this data, in \KMb fit we form azimuthal
angle asymmetries from these cross sections, and finally in the KM10 model we
take these cross sections explicitly into account. All fits have $\chi^2/{\rm d.o.f.}
\approx 1$ and, especially, we describe with the \KM model the Hall A cross
sections, where  the DVCS amplitude  enhancement  at large $\xB$ arises from an
effective contribution that is pragmatically associated with $\widetilde{\cal
H}$ and   $\widetilde{\cal E}$.
The results are available as executable code, providing the photon
electroproduction cross section off unpolarized proton,
on \href{http://calculon.phy.pmf.unizg.hr/gpd/}{http://calculon.phy.pmf.unizg.hr/gpd/}. 

\begin{figure}[h]
\centering
\includegraphics[width=0.45\textwidth]{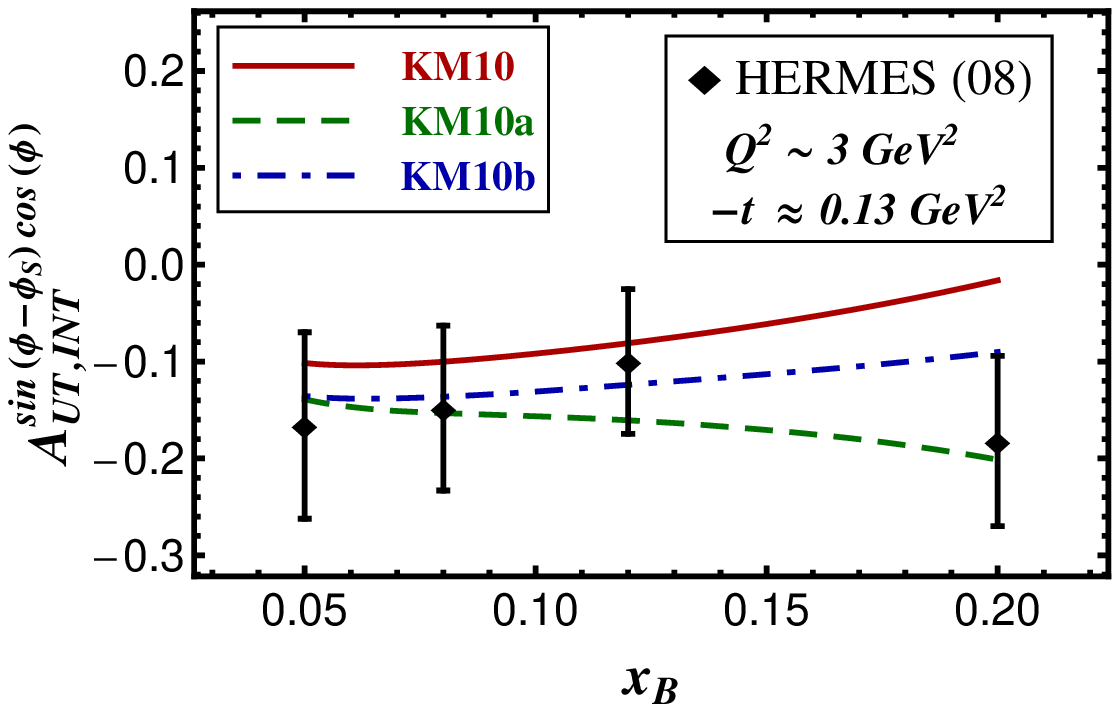}
\includegraphics[width=0.45\textwidth]{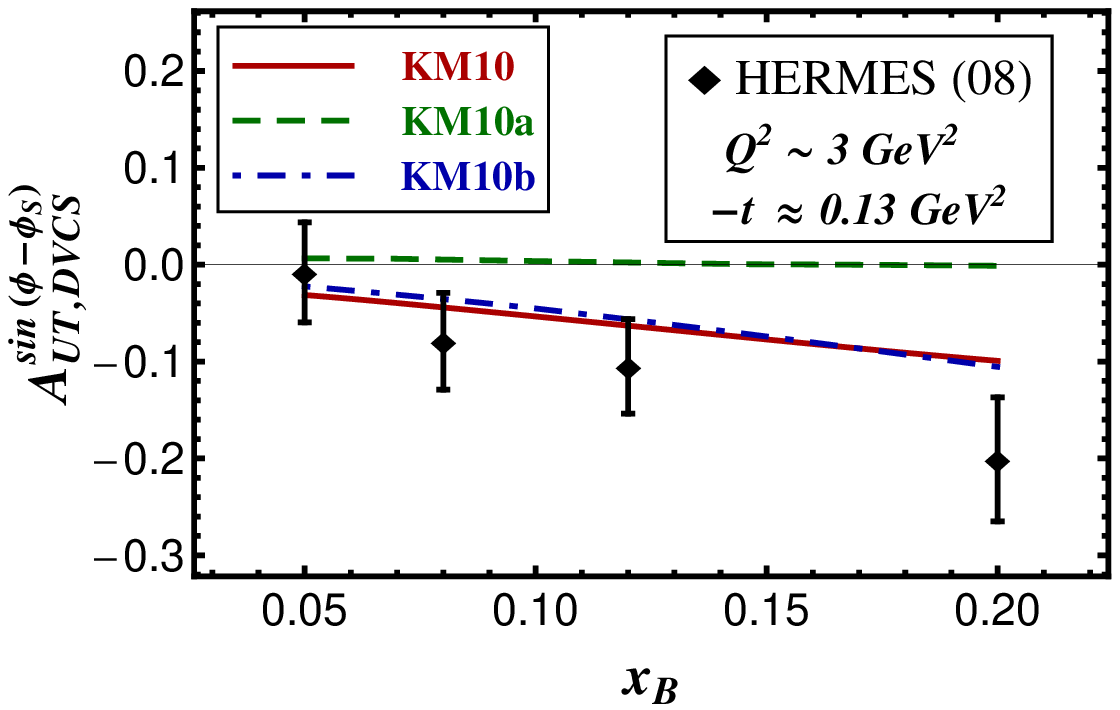}
\vspace{-4mm}
\caption{\small
DVCS predictions for the single transverse target spin asymmetries from HERMES \cite{Airapetian:2008aa}.
}
\label{Fig2}
\end{figure}
A more detailed model analysis of DVCS data, including measurements on polarized proton,  is feasible and should be performed in near future.
It has been illustrated in CFF fits \cite{Guidal:2010igGuidal:2010de} that the
inclusion of measurements with a longitudinal polarized proton target  provide
access to the GPD $\widetilde H$ and simultaneously reduce the GPD $H$
uncertainties. On the other hand, the target helicity flip GPDs $E$ and
$\widetilde E$ remain hidden in present measurements.  The most promising
observable to access GPD $E$ is the single transverse target spin asymmetry.
However, although the HERMES collaboration was able to disentangle the
interference and DVCS squared terms \cite{Airapetian:2008aa}, we might conclude
from the fact that
these measurements are partially describable with our unpolarized fits, where
GPD $E$ enters only in the real part of CFF $\cal E$, that  GPD $E$ extraction
from present data suffers strongly from the correlation with the other GPDs,
see Fig.~\ref{Fig2}.

\vspace{-2mm}
\section{Impact of planned and proposed DVCS experiments}

In nearest future it is expected that new DVCS measurements will be released:
beam spin asymmetry from HERMES, taken with a recoil detector, longitudinal
target spin asymmetries from CLAS, and cross sections from CLAS and Hall A, see
the contributions of M.~Murray and D.~Sokhan in these proceedings. These results
will provide more constraints in a global GPD analysis; however, they will not
give a full solution of the decomposition problem. Certainly,  the planned high
luminosity experiments at JLAB@12 GeV will help to access GPDs in the valence
region. The planned COMPASS II experiments will improve the
knowledge of GPD $H$ in the region where sea quarks and gluons are getting
dominant, and, hopefully, measurements on a polarized target might give insight
in the small $x$-behavior of other twist-two GPDs. Thereby, the 
interesting point is whether GPD $E$ possesses a ``pomeron''-like behavior.

\begin{figure}[t]
\centering
\vspace{-8mm}
\includegraphics[width=1\textwidth]{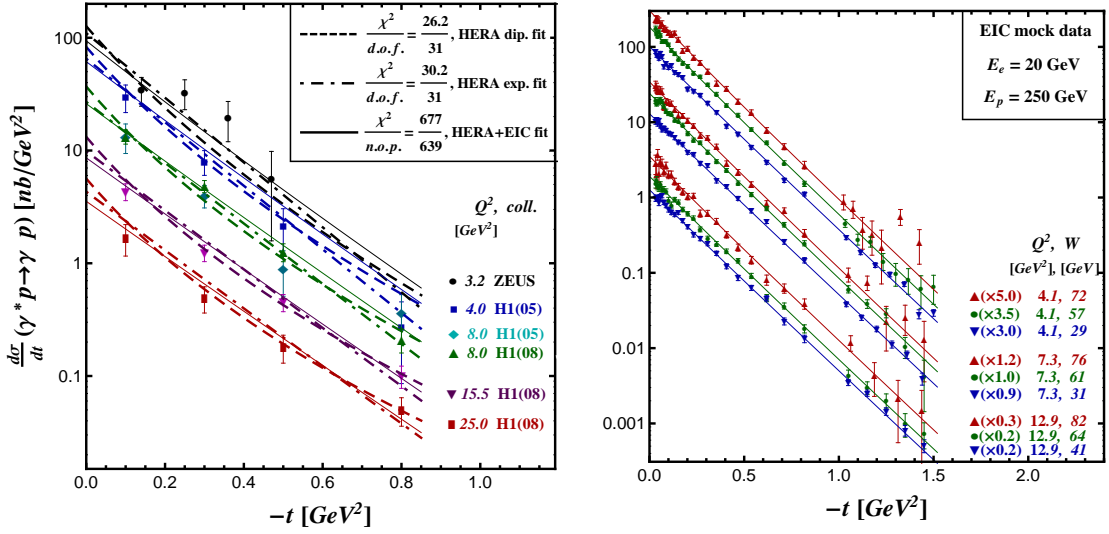}
\vspace{-8mm}
\caption{\small
DVCS measurements from H1  \cite{Aktas:2005tyAaretal09a} and ZEUS \cite{Chekanov:2003yaChekanov:2008vy} (left) and EIC pseudo data (right).
}
\label{Fig3}
\end{figure}
\begin{figure}[h]
\centering
\includegraphics[width=0.325\textwidth]{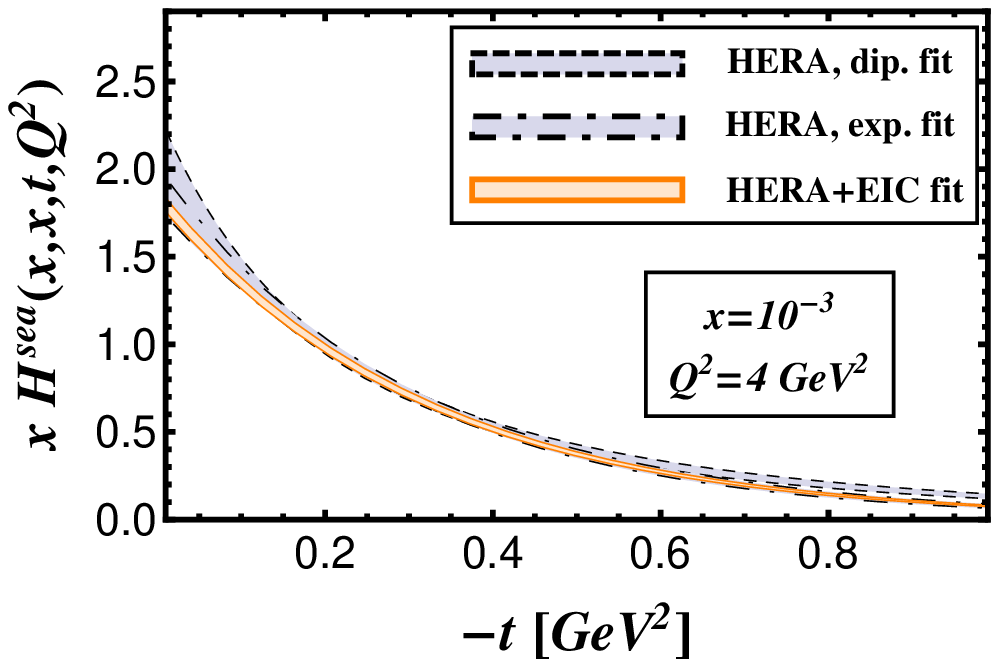}
\includegraphics[width=0.325\textwidth]{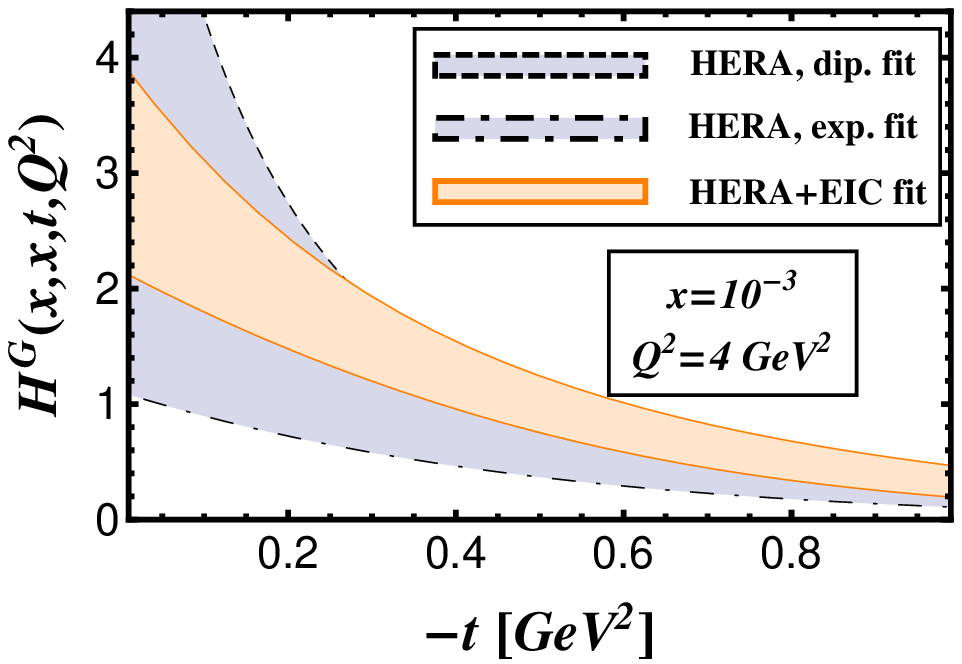}
\includegraphics[width=0.325\textwidth]{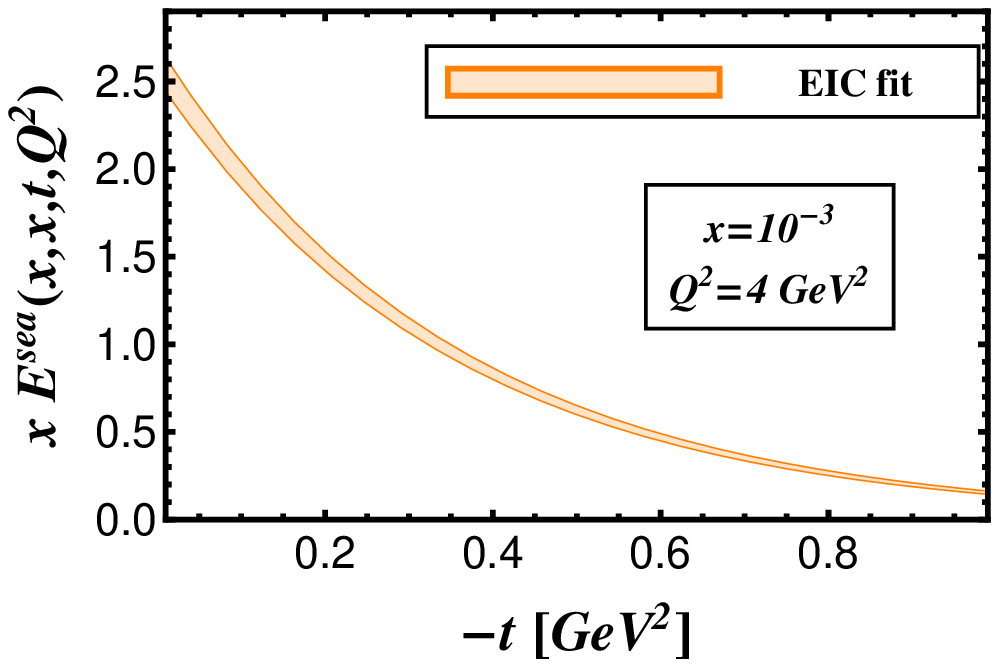}
\vspace{-2mm}
\caption{\small
Fit results to H1/ZEUS (dashed and dash-dotted surrounded areas) and EIC pseudo (solid surrounded areas)  DVCS cross sections, shown in
Fig.~\ref{Fig3}.
}
\label{Fig4}
\end{figure}
In Fig.~\ref{Fig3} we display DVCS cross section  data from H1/ZEUS (left) and
EIC pseudo data for an electron and proton  beam energy of $E_e=\unit{20}{GeV}$
and $E_p=\unit{250}{GeV}$, respectively (right).
The cross section is obtained by
subtracting the BH contribution, where the interference term is negligibly
small. Due to the exponential $t$-dependence of the CFFs, the subtraction
errors are rather large at $\unit{1}{GeV^2}\lesssim -t$.  For more details on
data simulation, see the contributions of E.~Aschenhauer and S.~Fazio in these
proceedings.
In Fig.~\ref{Fig4} we display the outcome of our H1/ZEUS (dashed and
dash-dotted surrounded areas) and simultaneous H1/ZEUS/EIC (solid surrounded
areas) fits with flexible GPD models. Thereby, we also took EIC pseudo data for
the transverse polarized target spin asymmetry $A_{\rm
UT}^{\sin(\phi-\phi_S)}(\phi)$, which allows for a decomposition of GPD $H$ and
$E$ contributions. As explained in the proceedings contribution of M.~Diehl,
such measurements allow for a 2D imaging of the partonic content at small $x$
of the unpolarized and transverse polarized proton. We add that an access of GPD $E$
in this region will provide a qualitative estimate of the angular
momentum carried by sea quarks.

\vspace{-2mm}
\section{Conclusions}
The first generation of hard exclusive experiments at HERA and JLAB provided us
insight into the GPD description of DVCS, where GPD $H$ could be accessed with
some uncertainty. The biggest portion of these uncertainties is related to the
fact that measurements on unpolarized proton do not allow for a GPD
decomposition. In future GPD analyses this problem might be partially
overcome, however, we expect that it cannot be fully solved. Planned and
proposed experiments will have a big impact to reveal GPDs from measurements.
Especially, a high-luminosity EIC offers the possibility to resolve the
transverse distribution of quarks and to give a qualitative insight into the
angular momentum of sea quarks.\\
\vspace{-2mm}

\noindent
{\bf Acknowledgements:} This work is
partly supported by MZOS grant no.~119-
0982930-1016,  BMBF grant no.~06BO9012 and
the 
HadronPhysics3 Grant Agreement no.~283286.

\noindent


\begin{thebibliography}{10}

\bibitem{Die03aBelRad05}
M.~Diehl.
\newblock Phys. Rept. {\bfseries 388} (2003) 41,
\href{http://arxiv.org/abs/hep-ph/0307382}{{\ttfamily hep-ph/0307382}};
A.~V. Belitsky and A.~V. Radyushkin.
\newblock Phys. Rept. {\bfseries 418} (2005) 1--387,
\href{http://arxiv.org/abs/hep-ph/0504030}{{\ttfamily hep-ph/0504030}}.

\bibitem{Airapetian:2009aa}
A.~Airapetian {\em et~al.}
\newblock \href{http://dx.doi.org/10.1088/1126-6708/2009/11/083}{JHEP
  {\bfseries 11} (2009) 083},
\href{http://arxiv.org/abs/0909.3587}{{\ttfamily arXiv:0909.3587 [hep-ex]}}.

\bibitem{Girod:2007aa}
F.~X. Girod {\em et~al.}
\newblock \href{http://dx.doi.org/10.1103/PhysRevLett.100.162002}{Phys. Rev.
  Lett. {\bfseries 100} (2008) 162002},
\href{http://arxiv.org/abs/0711.4805}{{\ttfamily arXiv:0711.4805 [hep-ex]}}.

\bibitem{Munoz_Camacho:2006hx}
C.~M. Camacho {\em et~al.}
\newblock \href{http://dx.doi.org/10.1103/PhysRevLett.97.262002}{Phys. Rev.
  Lett. {\bfseries 97} (2006) 262002},
\href{http://arxiv.org/abs/nucl-ex/0607029}{{\ttfamily arXiv:nucl-ex/0607029}}.



\bibitem{Kumericki:2009uqKumericki:2011zc}
K.~Kumeri{\v c}ki and D.~M{\"u}ller.
\newblock \href{http://dx.doi.org/10.1016/j.nuclphysb.2010.07.015}{Nucl. Phys.
  {\bfseries B841} (2010) 1--58},
\href{http://arxiv.org/abs/0904.0458}{{\ttfamily arXiv:0904.0458 [hep-ph]}};
K.~Kumeri{\v c}ki {\it et al}.
\newblock
\href{http://arxiv.org/abs/1105.0899}{{\ttfamily arXiv:1105.0899 [hep-ph]}}.


\bibitem{Goloskokov:2007nt}
S.~V. Goloskokov and P.~Kroll.
\newblock \href{http://dx.doi.org/10.1140/epjc/s10052-007-0466-5}{Eur. Phys. J.
  {\bfseries C53} (2008) 367--384},
\href{http://arxiv.org/abs/0708.3569}{{\ttfamily arXiv:0708.3569 [hep-ph]}}.

\bibitem{Guidal:2010igGuidal:2010de}
M.~Guidal.
\newblock \href{http://dx.doi.org/10.1016/j.physletb.2010.04.053}{Phys. Lett.
  {\bfseries B689} (2010) 156--162},
\href{http://arxiv.org/abs/1003.0307}{{\ttfamily arXiv:1003.0307 [hep-ph]}};
\newblock \href{http://dx.doi.org/10.1016/j.physletb.2010.07.059}{Phys. Lett.
  {\bfseries B693} (2010) 17--23},
\href{http://arxiv.org/abs/1005.4922}{{\ttfamily arXiv:1005.4922 [hep-ph]}}.

\bibitem{Airapetian:2008aa}
A.~Airapetian {\em et~al.}
\newblock \href{http://dx.doi.org/10.1088/1126-6708/2008/06/066}{JHEP
  {\bfseries 06} (2008) 066},
\href{http://arxiv.org/abs/0802.2499}{{\ttfamily arXiv:0802.2499 [hep-ex]}}.



\bibitem{Aktas:2005tyAaretal09a}
A.~Aktas {\em et~al.}
\newblock \href{http://dx.doi.org/10.1140/epjc/s2005-02345-3}{Eur. Phys. J.
  {\bfseries C44} (2005) 1--11},
\href{http://arxiv.org/abs/hep-ex/0505061}{{\ttfamily arXiv:hep-ex/0505061}};
F.~Aaron {\em et~al.}
\newblock \href{http://dx.doi.org/10.1016/j.physletb.2009.10.035}{Phys.Lett.
  {\bfseries B681} (2009) 391--399},
  \href{http://arxiv.org/abs/0907.5289}{{\ttfamily arXiv:0907.5289 [hep-ex]}}.

\bibitem{Chekanov:2003yaChekanov:2008vy}
S.~Chekanov {\em et~al.}
\newblock \href{http://dx.doi.org/10.1016/j.physletb.2003.08.048}{Phys. Lett.
  {\bfseries B573} (2003) 46--62},
\href{http://arxiv.org/abs/hep-ex/0305028}{{\ttfamily arXiv:hep-ex/0305028}};
\newblock \href{http://dx.doi.org/10.1088/1126-6708/2009/05/108}{JHEP
  {\bfseries 05} (2009) 108},
\href{http://arxiv.org/abs/0812.2517}{{\ttfamily arXiv:0812.2517 [hep-ex]}}.

\end{thebibliography}
\vspace{-3mm}
\providecommand{\href}[2]{#2}\begingroup\raggedright\endgroup

\end{document}